\documentstyle[preprint,aps,eqsecnum]{revtex}
\begin{document}
\draft
\preprint{Alberta Thy-52-93, hep-th/9411237}
\title{Discrete spectral shift in an anisotropic universe}
\author{D.\ J. Lamb\cite{email}, A.\ Z. Capri and
M. Kobayashi\cite{address} }
\address{ Theoretical Physics Institute, \\
Department of Physics,
University of Alberta,\\
Edmonton, Alberta T6G 2J1, Canada}
\date{\today}
\maketitle
\begin{abstract}
 In this paper we calculate the particle creation as seen by a stationary
observer in an anisotropic universe. By using an observer and geometry
dependent time to quantise a massive scalar field we  show that  a discrete
energy spectrum shift occurs. The length scale associated with the geometry
provides the energy scale by which the spectrum is shifted. The $\beta (p,q)$
coefficient for the Bogolubov transformation calculated is proportional to a
series of delta functions whose argument contains $p$ and $q$ and half
multiples of the root of the curvature.
\end{abstract}


\section{Introduction}
The problems surrounding  the construction of quantum fields in curved
spacetime backgrounds have seen renewed interest \cite{recent}. In particular
much of this interest is centered around attempts to define particles in curved
backgrounds and also various means of normal ordering expectation values to
obtain physically reasonable finite results. This is in contrast to much of the
earlier work which involved calculations of detector response functions or
elaborate means of regularizing and renormalizing expectation values of the
stress tensor.
 Of course these two problems are intimately related, if there are regions
where one can construct a Fock space, and thus have a particle interpretation,
a natural  normal ordering or vacuum subtraction procedure exists. The
difference in the recent literature is that the regions where one expects to be
able to do this are no longer required to be flat.

 The approach we use in this paper is that of Capri and Roy \cite{Caproy92}.
This is a coordinate independent geometrical approach where the geometry
determines a preferred time with which to quantize the field. This preferred
direction of time is along
a normal to a spacelike surface consisting of those spacelike geodesics which
are orthogonal to the observers 4-velocity. In this way the construction
depends only on the geometry, the observers position, and the tangent to the
observer's worldline. This approach is very similar to that of
\cite{eth} where they also construct a spacelike surface which is geodesic.
The major difference is that here a local Poincar\'e invariance is also used.

\section{The Model}
 The model  we investigate in this paper is an anisotropic $3+1$ generalization
of $1+1$ de Sitter space of constant curvature. The simplest generalization
being just the addition of a 2-plane. Specifically we are investigating
particle creation due to the gravitational field which is described by the
metric,
\begin{equation}
ds^2 = dT^2 - e^{\lambda T}(dX^1)^2 - (dX^2)^2 -(dX^3)^2.
\label{2.1}
\end{equation}
More precisely we investigate the particle creation as observed by an  observer
 stationary with respect to the coordinates $(T,X^1,X^2,X^3)$

 To follow the prescription of \cite{Caproy92} we first find the geodesics
 in this spacetime. The first integrals of the geodesics are:
 \begin{equation}
\frac{dX^1}{ds}= \frac{c_1}{e^{\lambda T}}\ , \ \ \ \frac{dX^2}{ds}= c_2\ , \ \
\ \frac{dX_3}{ds}= c_3\ , \ \ \ \ \frac{dT}{ds}=\sqrt{\epsilon
+\frac{c_1^2}{e^{\lambda T}}+c_2^2 + c_3^2 }
 \label{2.2}
\end{equation}
where $\epsilon = \pm 1$ depending on whether the geodesic is timelike or
spacelike respectively.

The preferred coordinates on the surface are constructed using a 4-bein of
orthogonal basis vectors at $P_0$, the observers position. We choose these
vectors to be,
\begin{equation}
e_0(P_0)=(1,0,0,0)\ \ \ e_1(P_0)=(0,e^{-\lambda\frac{T_0}{2}},0,0)\ \ \
e_2(P_0)=(0,0,1,0)\ \ \ e_3(P_0)=(0,0,0,1).
\label{2.4}
\end{equation}
In this way the tangent to the chosen observer's worldline at $P_0$ corresponds
to $e_0(P_0)$.

To construct the spacelike surface orthogonal to the tangent of the observer's
worldline we therefore require that,
\begin{equation}
\frac{dT}{ds}\left|_{P_0}=0 \right.\ \ \ {\rm which\ \ implies}\ \ \
\frac{c_1^2}{e^{\lambda T_0}}+c_2^2+c_3^2 = 1
\label{2.5}
\end{equation}
The preferred coordinates on the spacelike hypersurface are chosen to be
Riemann coordinates based on the observer's position
$P_0=(T_0,X^1_0,X^2_0,X^3_0)$. With $p^{\mu}$ given by the tangent vector, at
$P_0$, to the geodesic connecting $P_0$ to $P_1$. The point $P_1$ is the point
at which the timelike geodesic ``dropped" from an arbitrary point
$P=(T,X^1,X^2,X^3)$ intersects the spacelike surface orthogonally. The Riemann
coordinates $\eta^{\alpha}$ of the point $P_1$ are given by,
\begin{equation}
s p^{\mu} = \eta^{\alpha}e_{\alpha}^{\mu}(P_0)
\label{2.6}
\end{equation}
where $s$ is the distance along the geodesic $P_0-P_1$.
 Using $e_{\alpha}^{\mu}e_{\beta \mu} =
 \eta_{\alpha \beta}$ (Minkowski metric),
  and the orthogonality of $p^{\mu}$ to $e_0(P_0)$
  we have
\begin{equation}
\eta^0=sp^{\mu}e^0_{\mu}(P_0) \ \ \ \ \
 \eta^i=-sp^{\mu}e^i_{\mu}(P_0).
\label{2.7}
\end{equation}
The surface $S_0$ is just the surface $\eta^0=0$
and the coordinates $x^i$ are
\begin{equation}
x^1=s\frac{c_1}{\sqrt{e^{\lambda T_0}}}\ \ \ x^2=sc_2 \ \ \ x^3=sc_3
\label{2.8}
\end{equation}
where $s$ is the geodesic distance between the points $P_0$ and $P_1$.

The direction of time is given by the normal to the spacelike hypersurface and
the preferred time $t$ for the arbitrary point $P$ is given by the proper
distance along this timelike geodesic connecting $P$ to $P_1$. The timelike
geodesic is also determined by (\ref{2.2}) except with $\epsilon=-1$ and a
different choice of the constants which we denote by $b_i$. The condition that
the geodesic connecting $P$ to $P_1$ is normal to the spacelike hypersurface
requires that
\begin{equation}
\sqrt{\left(1+\frac{b_1^2}{e^{\lambda T_1}} + b_2^2 + b_3^2 \right)}\sqrt{
\left(\frac{c_1^2}{e^{\lambda T_1}} - \frac{c_1^2}{e^{\lambda T_0}} \right)} =
\frac{b_1c_1}{e^{\lambda T_1}} + b_2c_2 + b_3c_3.
\label{2.9}
\end{equation}
We can now calculate the dependence of $(T,X^1,X^2,X^3)$ on the preferred
coordinates $(t,x^1,x^2,x^3)$ and then calculate the metric in its preferred
form. To calculate this dependence we must use the above equations for $x^i$
(\ref{2.8}) and also calculate the change in the coordinates $X^i$ along the
spacelike and timelike geodesics which ultimately connect $P_0$ to $P$.
\begin{eqnarray}
X^1 &=& X^1_0 + \int_{T_0}^{T_1}dT\frac{c_1}{e^{\lambda T_1}}
\left(\frac{c_1^2}{e^{\lambda T}}- \frac{c_1^2}{e^{\lambda
T_0}}\right)^{-\frac{1}{2}} +  \int_{T_1}^{T}dT'\frac{b_1}{e^{\lambda
T'}}\left(1+\frac{b_1^2}{e^{\lambda T'}}+b_2^2+b_3^2 \right)^{-\frac{1}{2}}
\nonumber \\
X^2 &=& X^2_0 + \int_{T_0}^{T_1}dT\frac{c_2}{e^{\lambda T_1}}
\left(\frac{c_1^2}{e^{\lambda T}}- \frac{c_1^2}{e^{\lambda
T_0}}\right)^{-\frac{1}{2}} +  \int_{T_1}^{T}dT'\frac{b_2}{e^{\lambda
T'}}\left(1+\frac{b_1^2}{e^{\lambda T'}}+b_2^2+b_3^2 \right)^{-\frac{1}{2}}
\nonumber \\
X^3 &=& X^3_0 + \int_{T_0}^{T_1}dT\frac{c_3}{e^{\lambda T_1}}
\left(\frac{c_1^2}{e^{\lambda T}}- \frac{c_1^2}{e^{\lambda
T_0}}\right)^{-\frac{1}{2}} +  \int_{T_1}^{T}dT'\frac{b_3}{e^{\lambda
T'}}\left(1+\frac{b_1^2}{e^{\lambda T'}}+b_2^2+b_3^2 \right)^{-\frac{1}{2}}
\nonumber \\
\label{2.10}
\end{eqnarray}
and $t$ the proper distance along $P-P_1$
\begin{equation}
t=\int_{T_1}^{T}dT'\left(1+\frac{b_1^2}{e^{\lambda T'}}+b_2^2+b_3^2
\right)^{-\frac{1}{2}}.
\label{2.11}
\end{equation}
At this point we can see that if we choose $b_2=b_3=0$ this just corresponds
to aligning the spacelike hypersurfaces so that $X^2=X^2_1$ and $X^3=X^3_1$.
This simplifies the analysis considerably and gives the expected result that,
\begin{equation}
X^2=X^2_0 + x^2 \ \ \ \ {\rm and}\ \ \ \ X^3=X^3_0 + x^3.
\label{2.13}
\end{equation}
The only non-trivial part of the transformation therefore involves $(T,X^1)$
and $(t,x^1)$. By performing the above integral for $X^1$ and inverting the $t$
integral one is left with the coordinate transformations
\begin{eqnarray}
e^{\frac{\lambda}{2}(T-T_0)}&=&\sinh(\frac{\lambda t}{2})+\cosh(\frac{\lambda
t}{2})\cos(\frac{\lambda x^1}{2}) \nonumber \\
\frac{\lambda}{2}(X^1-X_0^1)e^{\lambda\frac{T}{2}} &=& -\cosh(\frac{\lambda
t}{2})\sin({\frac{\lambda x^1}{2}}).
\label{2.14}
\end{eqnarray}
In terms of the preferred coordinates $(t,x^i)$ the metric now has the form,
\begin{equation}
ds^2=dt^2-\cosh^2(\frac{\lambda t}{2})(dx^1)^2-(dx^2)^2-(dx^3)^2.
\label{2.15}
\end{equation}

This result is of course not a surprise to anyone familiar with the different
forms of de Sitter space in $1+1$ dimensions. Unfortunately the usual analysis
does not deal with the observer dependent nature of the coordinate
transformations. We will see that this is in fact where the interesting physics
comes from. Indeed if one proceeds to quantize the field on $t=$constant
surfaces it is easy to see that all these surfaces can be made to look like
Minkowski space. The point is that they cannot be made to all look like
Minkowski space simultaneously. It would therefore seem obvious that the
physics is going to be determined not by the form of the metric on a particular
surface but by the transformations relating one surface's preferred coordinates
to another surface's preferred coordinates.

\section{Modes and Quantization}
In the coordinates constructed in the last section the non-minimally coupled
massive Klein Gordon equation is
\begin{equation}
\partial_t^2\phi+\frac{1}{\sqrt{g}}\partial_t(\sqrt{g})\partial_t\phi
+\frac{1}{\sqrt{g}}\partial_i(\sqrt{g}g^{ij})\partial_j\phi+(m^2+\xi R)\phi=0
\label{3.0}
\end{equation}
This equation is strictly hyperbolic so long as $g^{ij}$ does not change sign.
The solutions are therefore uniquely determined by the initial data.

  To quantize a scalar field on the $t=0$ surface we now define the positive
frequency modes in the neighbourhood of this surface. The positive frequency
modes are defined as those which satisfy the initial conditions,
\begin{equation}
\phi_k^{+}(t,{\bf x})\left|_{t=0}\right.=A_k(0,{\bf x})\ \ \ {\rm and }\ \ \
\partial_t(\phi_k^{+}(t,{\bf x}))\left|_{t=0}\right.=-i\omega_k(0) A_k(0,{\bf
x})
\label{3.1}
\end{equation}
Where $A_k(t,{\bf x})$ are the instantaneous eigenmodes of the spatial part of
the Laplace-Beltrami operator, and $\omega_k(t)^2$ are the corresponding
eigenvalues.
\begin{equation}
\left[\frac{1}{\sqrt{g}}\partial_i\left(\sqrt{g}g^{ij}\partial_j \right)
+m^2+\xi R\right]A_k(t,{\bf x})=\omega_k^2(t)A_k(t,{\bf x}).
\label{3.2}
\end{equation}
Henceforth we just write $\omega_k$ for $\omega_k(0)$.
Due to the simple form of $g_{\mu\nu}$ at $t=0$ the eigenmodes and values take
on the simple form,
\begin{eqnarray}
A_k(0,{\bf x})&=&e^{i{\bf k \cdot x}} \\ \nonumber
\omega_k^2(0)&=&{\bf k}^2 + m^2 +\xi R.
\label{3.3}
\end{eqnarray}
Near the surface $t=0$ the second term of (\ref{3.0}) vanishes to $O(t^2)$,
this implies that the initial conditions for the time dependence of the field
which are also good to $O(t^2)$.

To impose these initial conditions we must find a complete set of modes for the
entire wave operator. Because the differential equation is separable we look
for solutions of the form $f_k(t)e^{i{\bf k \cdot x}}$. The differential
equation satisfied by the $f_k(t)$ is then,
\begin{equation}
\partial_t^2 f_k(t) + \frac{\lambda}{2} \tanh(\frac{\lambda t}{2})\partial_t
f_k(t) + \left(k_1^2 {\rm sech}^2(\frac{\lambda t}{2}) +k_2^2 +k_3^2 + m^2 +
\xi R \right)f_k(t)=0
\label{3.4}
\end{equation}
The positive frequency modes are those whose ``time" part satisfy the above
differential equation and the initial conditions
\begin{equation}
f_k(0)=1 \ \ \ {\rm and }\ \ \ \ \dot{f}_k (0) = -i\omega_k.
\label{3.5}
\end{equation}
The positive frequency modes are given in terms of hypergeometric functions
$H(a,b,c,x)$ by
\begin{equation}
\phi_k^+(t,{\bf x})=e^{i{\bf k \cdot x}}{\rm sech}(\frac{\lambda
t}{2})^{2n}\left\{\right.
H(\alpha,\beta,\frac{1}{2},\tanh^2(\frac{\lambda t}{2}))
-\left. i\frac{2\omega_k}{\lambda}
\tanh(\frac{\lambda
t}{2})H(\alpha+\frac{1}{2},\beta+\frac{1}{2},\frac{3}{2},\tanh^2(\frac{\lambda
t}{2}))\right\}
\label{3.6}
\end{equation}
where
\begin{eqnarray}
 n  &=& \frac{1}{4}-\frac{i}{\lambda}\sqrt{k_2^2+k_3^3+m^2+ \xi
R-\frac{\lambda^2}{16}} \nonumber \\
\alpha &=&
-\frac{k_1}{\lambda}+\frac{1}{4}-\frac{i}{\lambda}\sqrt{k_2^2+k_3^2+m^2+\xi R
-\frac{\lambda^2}{16}} \nonumber \\
\beta  &=&
\frac{k_1}{\lambda}+\frac{1}{4}-\frac{i}{\lambda}\sqrt{k_2^2+k_3^2+m^2+\xi R
-\frac{\lambda^2}{16}}. \nonumber \\
\label{3.7}
\end{eqnarray}

We can now write out the field which has been quantized on surface $1$ as,
\begin{equation}
\Psi_1=\int_{-\infty}^{\infty}dk \frac{1}{\sqrt{2\omega_k}}\left\{
\phi_k^+(t,{\bf x})a_1(k)+\phi_k^{+ \ast}(t,{\bf x})a_1^\dagger(k)\right\}
\label{3.8}
\end{equation}

\section{Particle Creation}
To investigate particle creation in the model universe as observed by an
observer stationary with respect to the original coordinates $(T,X^1,X^2,X^3)$
we calculate the Bogolubov transformation relating the annihilation and
creation operators from two different surfaces of quantization that the
observer passes through. To calculate the coefficients of this transformation
we equate the same field from two different quantizations on a common surface,
\begin{equation}
\Psi_1(t,x)=\Psi_2(t'(t,x),x'(t,x)).
\label{4.1}
\end{equation}
Here $\Psi_1(t,x)$ is the field written out explicitly in (\ref{3.8}) and
$\Psi_2(t',x')$ is the same field which has been quantized on a second surface
$t'=0$. The ``second" field is therefore quantized for the same observer as the
first but at some later time $T'_0$. At this time the remark made at the end of
the second section becomes clearer. All the physics of the observations made by
this observer are determined by the functions $t'(t,x)$, $x'(t,x)$ and the
derivatives of these functions with respect to $t$. In this way the geometry of
the spacetime via the coordinate independent prescription we have used,
determines the spectrum of created particles.

For simplicity we calculate the Bogolubov transformation by ``matching" the
field and its first derivative with respect to $t$ at $t=0$.
\begin{eqnarray}
a_1(k)&=&\frac{i}{(2\pi)^3}\frac{1}{\sqrt{2\omega_k}}\int d^3x
e^{i{\bf k \cdot x }}\left\{-i\omega_k\Psi_1(0,x)
+\left(\partial_t\Psi_1(t,x)\right)\left|_{t=0}\right. \right\}    \nonumber \\
   &=&\frac{i}{(2\pi)^3}\frac{1}{\sqrt{2\omega_k}}\int d^3x
e^{i{\bf k \cdot x }}\left\{-i\omega_k\Psi_2(t'(0,x),x'(0,x))
+\left(\partial_t\Psi_2(t'(t,x),x'(t,x))\right)\left|_{t=0}\right. \right\}
\nonumber \\
\label{4.2}
\end{eqnarray}
Using this equation, we can write out the Bogolubov transformation in the form
\begin{equation}
a_1(k)=\int d^3p \alpha(k,p)a_2(p) + \int d^3p \beta(k,p)a_2^{\dagger}(p).
\label{4.3}
\end{equation}
The spectrum of created particles is determined by $\left|\beta(k,p)
\right|^2$.
Writing out $\beta(k,p)$ explicitly we find it has some interesting properties
due to it's dependence on the inverse relations $t'(t,x)$,$x'(t,x)$,
\begin{equation}
\beta(k,p)=\frac{-i}{2\pi}\delta(p_{2}+k_{2})
\delta(p_{3}+k_{3})\int\! dx_1
\frac{e^{ik_1x_1}}{\sqrt{4\omega_p\omega_k}}\left\{ i\omega_k f_p^{+
\ast}(t'(0,x))e^{ip_1x'_1(0,x)}-\partial_t\left\{
f_p^{+ \ast}(t'(t,x))e^{ip_1x'_1(t,x)} \right\}\left|_{t=0}\right.
\right\}
\label{4.4}
\end{equation}
where
\begin{eqnarray}
x'(t,x)=\frac{2}{\lambda}\tan^{-1}\left(\frac{\cosh(\frac{\lambda t}{2})
\sin(\frac{\lambda x}{2})}{\cosh(\frac{\lambda t}{2})\cos(\frac{\lambda
x}{2})\cosh(\frac{\lambda}{2}(T'_0-T_0))-\sinh(\frac{\lambda
t}{2})\sinh(\frac{\lambda}{2}(T'_0-T_0))}
\right)  \nonumber \\
t'(t,x)=\frac{2}{\lambda}\sinh^{-1}\left(\sinh(\frac{\lambda
t}{2})\cosh(\frac{\lambda}{2}(T'_0-T_0))-\cosh(\frac{\lambda
t}{2})\cos(\frac{\lambda x}{2})\sinh(\frac{\lambda}{2}(T'_0-T_0)) \right).
\label{4.5}
\end{eqnarray}

\section{Discrete Shift of Energy Spectrum}
 Unfortunately due to the complicated nature of the expression for $\beta(k,p)$
we cannot write it out in a more transparent form which is still exact. We can
however discover some interesting facts about the spectrum of created particles
by investigating the integrand of the integral for $\beta(k,p)$.  In fact it is
not difficult to see that the particles observed by our stationary observer
possess a discrete energy spectrum shift. To see this we rewrite (\ref{4.4}) as
\begin{equation}
\beta(k,p)=\frac{-i}{2\pi}\int_{-\infty}^{\infty}dx \frac{e^{i(p_1+q_1)x_1}}{
\sqrt{4\omega_p\omega_k}}F(k,p,x)\delta(p_{2}+q_{2})\delta(p_{3}+q_{3})
\label{5.1}
\end{equation}
where
\begin{equation}
F(k,p,x)=e^{ip_1(x'(0,x)-x)}\left\{ i\omega_k f_p^{+\ast}(t'(0,x))
-iq_1\frac{\partial x'}{\partial t}f_p^{+\ast}(t'(t,x)) -\frac{\partial
t'}{\partial t}\partial_{t'}\left(  f_p^{+\ast}(t'(t,x))\right)
\right\}\left|_{t=0}\right.
\label{5.2}
\end{equation}
By inspection of the inverse relations (\ref{4.5}) one sees that $F(k,p,x)$
is a well behaved periodic function in $x$. The only difficulty arises with the
exponential factor. This factor is also periodic in $x$ if one is careful to
ensure that in the analysis both $x'$ and $x$ retain their range of $-\infty$
to $\infty$.

We can therefore write,
\begin{equation}
F(p,k,x)=\sum_{n=-\infty}^{\infty}C_n(p,k)e^{in\frac{\lambda x}{2}}
\label{5.3}
\end{equation}
which implies that,
\begin{equation}
\beta(p,k)=\frac{-i}{\sqrt{4\omega_p\omega_k}}\sum_{n=-\infty}^{\infty}C_n(p,k)
\delta(p_1+k_1+\frac{n\lambda}{2})\delta(p_{2}+k_{2})\delta(p_{3}+k_{3}).
\label{5.4}
\end{equation}
Unfortunately we cannot evaluate the $C_n(p,k)$ analytically but we can
evaluate them numerically for some specific values of $(T'_0-T_0)$,$\lambda$,
$\bf p$ and $\bf q$. This numerical analysis suggests that the particle
creation drops off rapidly for large $\bf p$ and $\bf q$. Nevertheless, it is
expected that the total particle creation, as in all such problems, is
infinite. The reason for this seems to be that the external field can pump in
an infinite amount of energy in a finite time \cite{sal}.
In this particular model the energy density of the classical matter field
giving rise to the geometry of the model is constant. If one calculates the
total energy of the classical matter field it is therefore infinite.

\section{Conclusions}
 We see from the above analysis that the particle creation due to the
gravitational field as seen by a stationary observer in the model universe, $
ds^2=dT^2-e^{\lambda T}(dX^1)^2-(dX^2)^2-(dX^3)^2$, observes a spectrum of
particles shifted by a  discrete amount. It appears that the one length scale
of the geometry namely $\sqrt{R}$ plays a role similar to the role the length
of a box plays for modes in a box. In this sense the discrete energy spectrum
shift may almost be expected.

\section{Acknowledgements}
This research was supported in part by a grant from
the Natural Sciences and Engineering Research Council
 of Canada.


\begin{references}
\bibitem[*]{email} Electronic Mail: lamb@phys.ualberta.ca
\bibitem[\dagger]{address} Permanent Address: Physics Department, Gifu
University, Yanagido, Gifu 501-11, Japan.
\bibitem{recent} B. Harms and Y. Leblanc UAHEP-939,hep-th/9308030 \\ C.M.
Massacand and C. Schmid, Annals of Physics (N.Y.), {\bf 231}, 363, (1994). \\
S. Massar ULB-TH 09/93, hep-th/9308085 \\
S.Massar, R.Parentani, R.Brout, ULB-TH-1/93, hep-th/9303147
\bibitem{Caproy92} A.Z. Capri and S.M. Roy, Modern
 Physics Letters A, {\bf 7}, 2317, (1992) also \\
 International Journal of Modern Physics A, {\bf 9}, 1239, (1994).
\bibitem{eth} second reference cited in \cite{recent}
\bibitem{sal} A.Salem and P.T. Mathews, Phys. Rev. {\bf 90}, 690, (1953).
\end{references}
\end{document}